\documentclass{optica-article}
\journal{oe}
\articletype{Research Article}

\usepackage{physics}
\newcommand{\Mod}[1]{\ (\mathrm{mod}\ #1)}
\DeclareMathOperator{\diag}{diag}
\DeclareMathOperator{\proj}{proj}
\DeclareMathOperator{\Normal}{Normal}
\DeclareMathOperator{\Uniform}{Uniform}

\usepackage{algorithm, algpseudocode}
\algnewcommand{\LeftComment}[1]{\State \(\triangleright\) #1}

\begin{document}

\title{Fast reconstruction of programmable integrated interferometers}

\author{
Boris Bantysh,\authormark{1,2,*}
Konstantin Katamadze,\authormark{1}
Andrey Chernyavskiy,\authormark{1,2} and
Yurii Bogdanov\authormark{1,2}
}

\address{
\authormark{1}Valiev Institute of Physics and Technology, Russian Academy of Sciences, 117218, Moscow, Russia,\\
\authormark{2}Russian Quantum Center, Skolkovo, Moscow 143025, Russia
}
\email{\authormark{*}bbantysh60000@gmail.com}

\begin{abstract}
Programmable linear optical interferometers are important for classical and quantum information technologies, as well as for building hardware-accelerated artificial neural networks. Recent results showed the possibility of constructing optical interferometers that could implement arbitrary transformations of input fields even in the case of high manufacturing errors. The building of detailed models of such devices drastically increases the efficiency of their practical use. The integral design of interferometers complicates its reconstruction since the internal elements are hard to address. This problem can be approached by using optimization algorithms [Opt. Express {\bfseries 29}, 38429 (2021)]. In this paper, we present a novel efficient algorithm based on linear algebra only, which does not use computationally expensive optimization procedures. We show that this approach makes it possible to perform fast and accurate characterization of high-dimensional programmable integrated interferometers. Moreover, the method provides access to the physical characteristics of individual interferometer layers.
\end{abstract}

\section{Introduction}
The use of classical and quantum properties of light opens up new possibilities for information processing methods \cite{minzioni2019roadmap}. Programmable optical circuits play an important role in classical and quantum optical communications \cite{carolan2015universal,harris2018linear,wang2018multidimensional,wang2020integrated}, in quantum computing \cite{schwartz2016deterministic,zhong201812,asavanant2019generation}, and in machine learning problems \cite{hamerly2019large,wetzstein2020inference,zhang2021optical}. The key element of such devices is a programmable linear optical interferometer. When control parameters are fixed, it performs a linear transformation of $N$ input fields:
\begin{equation}\label{eq:fields_transformation}
    \mqty(E_1^{\text{(out)}} & E_2^{\text{(out)}} & \cdots & E_N^{\text{(out)}})^T
    = V \cdot \mqty(E_1^{\text{(in)}} & E_2^{\text{(in)}} & \cdots & E_N^{\text{(in)}})^T,
\end{equation}
where $V$ is $N \times N$ complex transfer matrix of the interferometer. By changing the values of the control parameters one can change $V$. This dependence is determined by the interferometer internal structure, i.e. by the position of beam splitters (both two- and multichannel) and phase shifters \cite{reck1994experimental,clements2016optimal,fldzhyan2020optimal,saygin2020robust}. In the absence of losses, such structures are capable of implementing an arbitrary unitary transformation $V$ of the fields. Moreover, a proper choice of the architecture can achieve this even in the presence of high manufacturing errors \cite{fldzhyan2020optimal,saygin2020robust,pai2019matrix}.

One can tune the interferometer to some target unitary transformation $V$ in two ways. The first way is empirical. For each $V$, one optimizes the control parameters in order to bring the experimental results of optical fields measurements closer to the target ones \cite{hughes2018training}. This does not require building an exact model of interferometer internal structure, but is impractical, since it is necessary to measure all matrix elements at each step. For each new target matrix $V$, the procedure must be repeated. Another way is to first build a complete mathematical model of the interferometer based on the results of some set of measurements \cite{kuzmin2021architecture}. Then, the search for the desired set of control parameters is performed without any additional measurements. This procedure faces some difficulties. Since large-scale linear optical devices can be implemented in the integral design only, it is almost impossible to split the large-scale reconstruction problem into several low-scale ones. It needs additional optical outputs or modular optical circuits \cite{mennea2018modular} that causes additional losses and instability. Trying to characterize an integrated device as a whole, one has to face the fact that the results of each measurement are determined by a large number of internal device parameters. As a result, the efficiency of the optimization procedure rapidly decreases with increasing dimension $N$.

In this paper, we present a method for constructing a model of a programmable integrated interferometer based on algebraic operations only. This approach needs limited number of operations and is much more computationally efficient than the one using numerical optimization. The method is based on the layered interferometer architecture \cite{saygin2020robust}, which is described in Section~\ref{sect:model}. Further in Section~\ref{sect:reconstruction}, we describe the procedure for reconstructing interferometer mixing layers. As in \cite{kuzmin2021architecture}, it requires the measurement of all matrix elements of the complex transfer matrix $V$ (up to a global phase) for a given configuration of phase layers. Common methods for measuring the transfer matrices of optical interferometers using laser \cite{rahimi2013direct}, thermal \cite{katamadze2021linear}, and two-photon \cite{laing2012super,peruzzo2011multimode} fields make it possible to measure matrix elements up to input and output phases only. A non-ambiguous measurement of the matrix $V$ can be performed using homodyne detection of all the input and output coherent fields in \eqref{eq:fields_transformation} and requires $O(N^2)$ measurements \cite{jacob2018direct}. In Section \ref{sect:simulation}, we show that the proposed algebraic approach allows one to predict the transformation matrix $V$ for an arbitrary configuration of phase layers with an accuracy not lower than that for the method based on numerical optimization. The mixing layers are determined ambiguously, however, in the case of homogeneous losses, our method makes it possible to uniquely determine the normalized transmission coefficients of each individual layer. In addition, we show that the method is highly robust to phase layers control errors.

\section{Interferometer model}\label{sect:model}
\begin{figure}[htbp]
\centering\includegraphics[width=\linewidth]{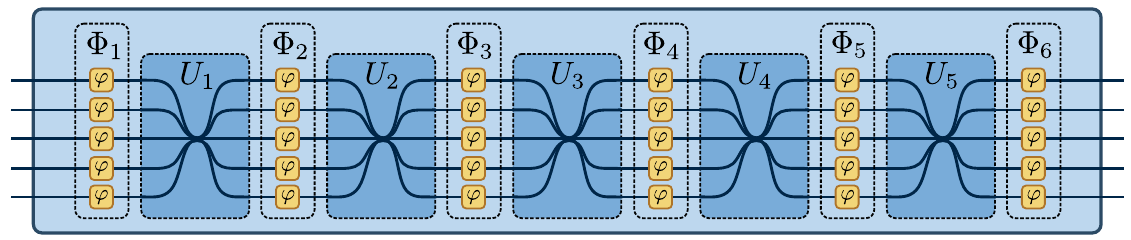}
\caption{Schematic representation of the layered architecture of a programmable interferometer. The number of mixing layers $K$ is equal to the number of channels $N=5$.}
\label{fig:interferometer}
\end{figure}

Below we consider the universal layered interferometer architecture consisting of alternating mixing and phase layers \cite{saygin2020robust,kuzmin2021architecture} (Fig.~\ref{fig:interferometer}). The corresponding transfer matrix has the form
\begin{equation}\label{eq:chip}
    V = \Phi_{K+1} U_K \Phi_K \dots U_2 \Phi_2 U_1 \Phi_1.
\end{equation}
Here $U_k$ and $\Phi_k$ are transfer matrices of the mixing and phase layers, respectively. Each mixing layer is an $N$-channel transformation that sufficiently distributes the signal of each input channel to all the output channels. All matrices $U_k$ can be different. Ideally, they are unitary, but this property is lost in the presence of linear losses. For the phase layer, each channel is subjected to a controlled phase shift through a thermo- or electro-optical effect. The corresponding transfer matrix has a diagonal form:
\begin{equation}
    \Phi_k = \diag\qty[\exp(i\varphi_n^{(k)})], \quad n = 1,\dots,N, \quad k = 1,\dots,K+1.
\end{equation}
Expression \eqref{eq:chip} can also be used to describe any other interferometer architecture (for example, the Reck \cite{reck1994experimental} or Clements \cite{clements2016optimal} architecture). In this case, $U_k$ can be the transfer matrices of beam splitters acting on two channels only, and the phase layers do not control all channels, but only a few ones.

With the number of mixing layers $K=N$, the layered architecture makes it possible to obtain an arbitrary transformation $V$ due to the proper choice of phases $\varphi_n^{(k)}$. We mostly consider this particular case.

The problem of constructing an interferometer model is reduced to determining all transfer matrices $U_k$ and $\Phi_k$. The dependence of $\varphi_n^{(k)}$ on the control parameters can be pre-calibrated even for unknown $U_k$. For this, a simple $\sin\varphi$ dependence of the interferometer output intensity on the value $\varphi$ of any single controlled phase is used (all other phases are disabled) \cite{carolan2015universal}. Below we assume that all phase layers are pre-calibrated (later we will show that our reconstruction method is robust to phase control errors). Reconstruction of the matrices $U_k$ is complicated by the fact that there is no way to ``turn off'' all mixing layers, except for one, and perform its measurement. However, in the next section, we show that it is possible to obtain a matrix equation that depends on a single mixing layer only by an appropriate choice of phase layers configurations. This equation is reduced to an eigendecomposition problem, which is solved algebraically.

\section{Mixing layers reconstruction}\label{sect:reconstruction}
In this section, we assume that we have full control over the phase layers. Later in Section~\ref{sect:simulation}, we show that the proposed method is robust to control errors. We also assume that none of the mixing layers has absolute losses in any channel, so all the matrices $U_1, \dots U_K$ are invertible. We also do not restrict them to be unitary.

We describe the reconstructing algorithm in several steps. First, we consider the ideal case when, for any configuration of the interferometer, we know exactly its full transfer matrix $V$. Next, we consequentially take into account the impossibility of experimental determination of the global phase of $V$, the presence of errors in measurements and phase layers control, and the special case of homogeneous losses inside the interferometer. Finally, we present the pseudo-code for the complete algorithm. Its software implementation is available online \cite{lib_github}.

\subsection{Ideal case}
Consider two configurations of the interferometer phase layers. The first configuration has all the phase layers disabled. The corresponding transfer matrix is $V_0=U_K\dots U_2U_1$. Next, the phases $\{\varphi_n, n=1,\dots,N\}$ are set in the second phase layer: $V_1=U_K\dots U_2\Phi U_1$, where $\Phi=\diag[\exp(i\varphi_n)]$. Consider the matrix $A_1=V_0^{-1}V_1=U_1^{-1}\Phi U_1$. Since the matrix $\Phi$ is known and diagonal, this equation is solved with respect to $U_1$ by the diagonalization of $A_1$. By induction, one can iteratively obtain all mixing matrices up to $U_{K-1}$. In particular, we enable only the $(k+1)$-th phase layer to obtain the $k$-th mixing layer:
\begin{equation}
    V_k = U_K \dots U_{k+1} \Phi U_k \dots U_2 U_1.
\end{equation}
Using $V_k$ and the results for all previous $k-1$ matrices we compute the following matrix:
\begin{equation}\label{eq:ak_mat}
    A_k = U_{k-1}\dots U_1 V_0^{-1} V_k U_1^{-1}\dots U_{k-1}^{-1} = U_k^{-1}\Phi U_k.
\end{equation}
The following equation determines its eigendecomposition:
\begin{equation}\label{eq:eigen}
    A_k\vec{u}_q = \lambda_q\vec{u}_q, \quad q=1,\dots,N.
\end{equation}
Since eigenvectors diagonalize $A_k$, we can use it to construct $U_k^{-1}$. To do this, one must sort the solutions of \eqref{eq:eigen} in such a way that matches eigenvalues with the diagonal of $\Phi$. Let us define the sorting indices $S=\{s_n: \lambda_{s_n}=\exp(i\varphi_n)\}$. Then the inverse transfer matrix has the form
\begin{equation}
    U_k^{-1} = \mqty(\vec{u}_{s_1} & \vec{u}_{s_2} & \cdots & \vec{u}_{s_N}).
\end{equation}
This procedure goes successively for $k=1,\dots,K-1$. Finally, the transfer matrix of the last mixing layer is $U_K=V_0U_1^{-1}U_2^{-1}\dots U_{K-1}^{-1}$. The procedure does not involve the first and last phase layers.

Note that for the correct work of the described method it is necessary that all $\lambda_q$ from \eqref{eq:eigen} (so the phases $\varphi_n$) must be different. Eigenvectors with equal eigenvalues cannot be uniquely defined, this leads to errors in constructing the columns of $U_k^{-1}$.

\subsection{Ambiguity}
\begin{figure}[htbp]
\centering\includegraphics[width=.9\linewidth]{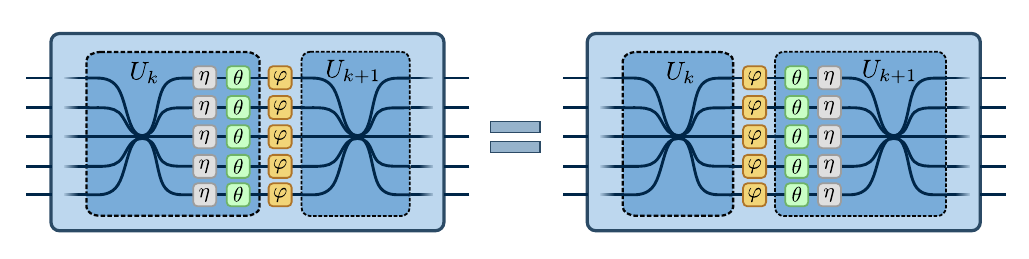}
\caption{Illustration of the impossibility to distinguish the losses and phase shifts at the output of one mixing layer from the losses and phase shifts at the input of the next mixing layer. Boxes with symbols $\theta$ and $\eta$ denote phase shifts and linear losses, respectively.}
\label{fig:ambiguity}
\end{figure}

The described algorithm determines all mixing layers of the interferometer up to following transformations:
\begin{equation}\label{eq:ambiguity}
    U_1 \rightarrow \Lambda_1 U_1, \quad
    U_k \rightarrow \Lambda_k U_k \Lambda_{k-1}^{-1} (1 < k < K), \quad
    U_K \rightarrow U_K \Lambda_{K-1}^{-1},
\end{equation}
where $\Lambda_k$ are arbitrary invertible diagonal matrices. Such ambiguity is caused by the fact that the equation \eqref{eq:eigen} is invariant under the multiplication of the eigenvector by an arbitrary number. This, however, is not the limitation of the proposed method, but the consequence of the impossibility to address each individual mixing layer separately in the experiment. Indeed, let us write the transfer matrix of the interferometer with arbitrary control phases taking \eqref{eq:ambiguity} into account:
\begin{equation}\label{eq:ambiguity2}
\begin{aligned}
    V
    &= \Phi_{K+1} (U_K \Lambda_{K-1}^{-1}) \Phi_K \dots (\Lambda_k U_k \Lambda_{k-1}^{-1}) \Phi_k \dots \Phi_2 (\Lambda_1 U_1) \Phi_1 \\
    &= \Phi_{K+1} U_K (\Lambda_{K-1}^{-1}\Lambda_{K-1}) \Phi_K \dots (\Lambda_k^{-1}\Lambda_k)\Phi_{k+1} U_k (\Lambda_{k-1}^{-1}\Lambda_{k-1}) \Phi_k \dots (\Lambda_1^{-1}\Lambda_1) \Phi_1 U_1 \\
    &= \Phi_{K+1} U_K\Phi_K \dots \Phi_{k+1} U_k \Phi_k \dots \Phi_2 U_1 \Phi_1,
\end{aligned}
\end{equation}
which is equivalent to the initial transfer matrix. We have used the commutativity of diagonal matrices $\Lambda_k$ and $\Phi_k$ in the second line of the equation \eqref{eq:ambiguity2}. Thus, transformations \eqref{eq:ambiguity} cannot be observed in the experiment, and they do not affect the ability of the model to predict the transfer matrix for any configuration of phase layers. Physically this means that we cannot distinguish the losses and constant phase shifts at the output of one mixing layer from the losses and phase shifts at the input of the next mixing layer (Fig.~\ref{fig:ambiguity}).

\subsection{Effect of the global phase}
\begin{figure}[htbp]
\centering\includegraphics[width=\linewidth]{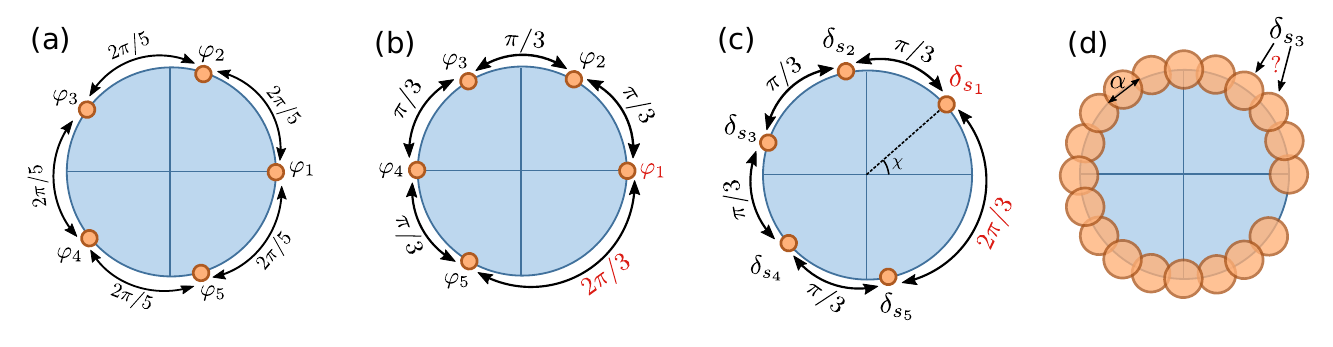}
\caption{The distribution of phases on the unit circle. (a) Uniform distribution of controllable phases on the single phase layer for $N = 5$. If the constant $\chi$ is a multiple of $2\pi/5$ then the transformation $\varphi_n \rightarrow \varphi_n+\chi\Mod{2\pi}$ leads to a similar distribution making it impossible to unambiguously compare the eigenvalues $\lambda_q=\exp(i\delta_q)$ with the phases $\varphi_n$. (b) The choice of asymmetric phase distribution \eqref{eq:phases_choice} distinguishes $\varphi_1$ from the others by the greatest distance to the previous phase. (c) The search for the largest distance between adjacent phases of eigenvalues $\lambda_q$ allows one to determine the index $q$ corresponding to the phase $\varphi_1$ and retrieve the constant $\chi$. (d) Strong phase fluctuations of the eigenvalues $\lambda_q$ at large $N$ (in this case $N = 19$) violates the sorting of the eigenvectors.}
\label{fig:phases}
\end{figure}

The approach described above gives an exact solution, when the matrices $V_k$ are exactly known. In practice, the elements of the transfer matrices are measured experimentally and, therefore, differ from the original ones. These differences are primarily related to the undetectable global phase. In this regard, in equation \eqref{eq:eigen}, we obtain eigenvalues $\lambda_q=\exp(i\delta_q)$, where the set of phases $\{\delta_q\}$ differs from the original set $\{\varphi_n\}$ by some constant value $\chi$. It makes the matching between $\{\delta_q\}$ and $\{\varphi_n\}$ ambiguous and could violate the sorting of eigenvectors. For example, a set of phases
\begin{equation}
    \varphi_n = \frac{2\pi(n-1)}{N}, \quad n=1,\dots,N
\end{equation}
from Fig.~\ref{fig:phases}a is invariant under $2\pi/N$ rotation and can't be used to determine $\chi$ in a unique way. This problem is solved by an appropriate choice of controllable phases such that $\{\varphi_n+\chi \Mod{2\pi}\} \neq \{\varphi_n\}$ for any non-trivial $\chi$. Consider the example presented in Fig.~\ref{fig:phases}b:
\begin{equation}\label{eq:phases_choice}
    \varphi_n = \frac{2\pi(n-1)}{N+1}, \quad n=1,\dots,N.
\end{equation}
One can observe that this set is invariant under the full $2\pi$ rotations only. Further, it is convenient to move from absolute phases to relative phases between adjacent channels of the phase layer (channel $N$ is considered adjacent to channel $1$):
\begin{equation}
    \Delta \varphi_n = \begin{cases}
        \varphi_1 - \varphi_N \Mod{2\pi}, & n = 1 \\
        \varphi_n - \varphi_{n-1} \Mod{2\pi}, & n > 1.
    \end{cases}
\end{equation}
It is easy to see that the relative phase for the first channel is twice the others for the choice \eqref{eq:phases_choice}:
\begin{equation}\label{eq:delta_phases}
    \Delta \varphi_1 = \frac{4\pi}{N+1}, \quad \Delta \varphi_{n>1} = \frac{2\pi}{N+1}.
\end{equation}
We can then calculate the sorting indices of eigenvalues from \eqref{eq:eigen} by relative phases that are insensitive to the global phase $\chi$:
\begin{equation}
    S = \qty{s_n: \Delta\delta_{s_1}=\Delta\varphi_1;\; \delta_{s_{n>1}}=\delta_{s_{n-1}}+\Delta\varphi_n}.
\end{equation}

Systematic and statistical errors also distort the spectrum of the matrix $A_k$, leading to more general eigenvalues: $\lambda_q=r_q\exp(i\delta_q)$, where $r_q\approx1$, $\delta_q\approx\varphi_q+\chi$. If the errors are not too large then we can again use the above approach, but instead of an exact equality we look for an approximate one:
\begin{equation}\label{eq:sorting_rule}
    S = \qty{
    s_n:
    s_1 = \operatorname*{arg\,max}_q \Delta\delta_q;\;
    s_{n>1} = \operatorname*{arg\,min}_q \qty[\delta_q - \delta_{s_{n-1}} \Mod{2\pi}]
    }.
\end{equation}
Note that in practice this minimization is not necessary. The set \eqref{eq:phases_choice} is sorted in ascending order, so one could also sort all $\delta_q$ and then find the maximum value of $\Delta\delta_q$. In view of \eqref{eq:delta_phases}, this value corresponds to the index $s_1$. The remaining elements will follow it on the unit circle in the anti-clockwise direction (Fig.~\ref{fig:phases}c).

\subsection{Noise handling}\label{sect:noise_handling}
\begin{figure}[htbp]
\centering\includegraphics[width=\linewidth]{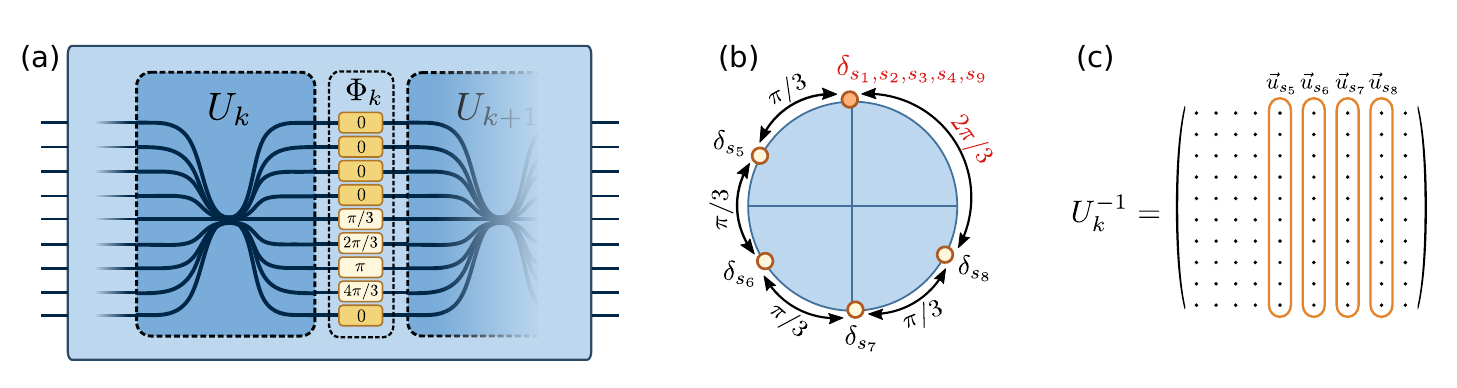}
\caption{An example of a partial reconstruction of the inverse matrix of the $k$-th mixing layer. The dimension of the interferometer is $N=9$. $M = 4$ matrix columns from 5 to 8 are estimated. (a) The $k$-th phase layer configuration. (b) The phases of matrix $A_k$ eigenvalues are sorted on the unit circle in ascending order. The first element is considered to be the one that is separated from the previous one by the largest distance. The first $N-M=5$ elements correspond to eigenvectors with indices $n\notin C$. The remaining $M = 4$ elements are eigenvectors with indices $n\in C$. (c) The corresponding eigenvectors are set to be the columns of the matrix $U_k^{-1}$.}
\label{fig:partial}
\end{figure}

The described algorithm works well on small dimensions. However, as $N$ increases, the density of $\{\varphi_n\}$ values on the unit circle increases. As a result, the eigenvalue sorting rule becomes more noise sensitive (Fig.~\ref{fig:phases}d). One can reduce this effect by activating only a few phases at a time and calculating only a few columns of the matrix $U_k^{-1}$. Denote by $C=\{c_m\}$ the set of indices of these columns and select the following set of phases (Fig.~\ref{fig:partial}a):
\begin{equation}\label{eq:phases_partial}
    \varphi_n = \begin{cases}
        0, & n \notin C, \\
        \frac{2\pi m}{M+2}, & n=c_m,
    \end{cases} \quad
    n = 1,\dots,N,
\end{equation}
where $M < N$ is the size of $C$. The set \eqref{eq:phases_partial} contains $M+1$ unique values. After measuring the interferometer transfer matrix $V_k$ and obtaining the matrix $A_k$ from \eqref{eq:ak_mat}, its eigenvalues are sorted according to rule \eqref{eq:sorting_rule}. It turns out that the values $\delta_{s_n}$ for $1\leq n \leq N-M$ are close to each other, and $\Delta \delta_{s_1}$ is maximal (Fig.~\ref{fig:partial}b). The remaining $M$ values correspond to the desired eigenvectors (Fig.~\ref{fig:partial}c).

The smaller the number $M$ of simultaneously estimated columns of $U_k^{-1}$, the less densely the corresponding phase values can be located. As a result it is possible to resolve the individual eigenvalues of the matrix $A_k$ from \eqref{eq:eigen} quite well even for large values of dimension $N$. In the limit $M=1$, we reconstruct the columns of the matrix $U_k^{-1}$ one by one. This, however, complicates the experiment: it is required to measure $\propto N/M$ transfer matrices for each mixing layer. The total number of transfer matrices used is $L=(K-1)\lceil N/M \rceil + 1$.

Note that in the above algorithm, the phases $\delta_q$ of $A_k$ eigenvalues were used for the correct eigenvectors sorting only. At the same time, the absolute values of $\delta_q$ contain information about the true values of the controlled phases $\varphi_n$ at the phase layer. These values can serve for more accurate calibration of the phase layer, taking its cross-talk into account.

\subsection{The case of homogeneous losses}\label{sect:no_losses}
The described procedure does not impose any constraints on resulting matrices $U_k$. In certain cases, it may be known \textit{a priory} that the interferometer losses are homogeneous. Then each $U_k$ belongs to the set of unitary matrices up to a constant factor $\eta_0$ describing the homogeneous losses. Thus, one can enhance the accuracy by projecting each $U_k$ to this set. For some non-unitary matrix $X$, the projection is given by
\begin{equation}\label{eq:uni_projection}
    \proj[X] = QW, \quad X = QSW^\dagger.
\end{equation}
Here matrices $Q$, $S$ and $W$ form the singular value decomposition of $X$. The matrices $Q$ and $W$ are unitary, while $S$ is non-negative diagonal matrix and equals to the identity matrix for unitary $X$. Within the unitary model, the transformations $\Lambda_k$ from \eqref{eq:ambiguity} are limited by arbitrary phase shifts only (there are no linear loss blocks in Fig.~\ref{fig:ambiguity}). This ambiguity does not affect the normalized transmission coefficients of mixing layers
\begin{equation}\label{eq:transmission}
    [T_k]_{mn} = \frac{\abs{[U_k]_{mn}}^2}{\frac{1}{N}\Tr(U_k^\dagger U_k)} = \frac{1}{\eta_0^2} \abs{[U_k]_{mn}}^2.
\end{equation}
Thus, this coefficients are defined unambiguously. Below we consider the case $\eta_0=1$ since homogeneous losses could be measured separately and compensated by collecting more data.

\subsection{Algorithm pseudo-code}\label{sect:pseudo_code}
Algorithms~\ref{alg:cap} summarizes the described procedure of mixing layers reconstructions. It takes the following parameters as the input: $N$ -- the number of input and output channels of the interferometer; $K$ -- the number of mixing layers; $M$ -- the maximal number of $U_k^{-1}$ columns, estimated at once. The output of the algorithm are transfer matrices $U_k$ ($k=1,\dots,K$) of all mixing layers.

The algorithm software implementation is available online \cite{lib_github}.

\begin{algorithm}[ht]
\caption{Mixing layers reconstruction}\label{alg:cap}
\begin{algorithmic}
\State Measure $V_0$ (all phase layers are disabled)
\State $V_T \gets I$ \Comment{The product of all currently estimated $U_k$}
\For{$k=1, \dots, K-1$}
    \State $U_k^{(\text{inv})} \gets I$ \Comment{Initial estimation}
    \State $j \gets 0$ \Comment{Current number of estimated columns}
    \While{$j < N$}
        \State $M^\prime = \text{min}(M, N-j)$
        \If{$M^\prime = N$} \Comment{All columns of $U_k^{-1}$ are estimated at once}
            \State $\varphi_n \gets \frac{2\pi(n-1)}{N + 1}$ ($n = 1, \dots, N$)
        \Else \Comment{Only $M^\prime$ columns of $U_k^{-1}$ are estimated}
            \State $\varphi_n \gets 0$ ($n = 1, \dots, N$)
            \State $\varphi_{j + m} \gets \frac{2\pi m}{M^\prime + 2}$ ($m = 1,\dots,M^\prime$)
        \EndIf
        \State Measure $V_k$ (phases of $(k+1)$-th phase layer are $\varphi_n$)
        \State $A_k \gets V_T V_0^{-1} V_k V_T^{-1}$
        \State Solve $A_k \vec{u}_q = r_q e^{i\delta_q} \vec{u}_q$ for $\vec{u}_q$, $r_q \geq 0$, $\delta_q \in [0, 2\pi)$ ($q = 1, \dots, N$)
        \State Find sorting indices $S$ using \eqref{eq:sorting_rule}
        \State $[U_k^{(\text{inv})}]_{n, j+m} \gets [\vec{u}_{s_{N-M^\prime+m}}]_n$ ($m=1,\dots,M^\prime$) \Comment{Update $M^\prime$ columns of $U_k^{-1}$ to match sorted eigenvectors of $A_k$}
        \State $j \gets j + M^\prime$
    \EndWhile
    \State $U_k \gets [U_k^{(\text{inv})}]^{-1}$
    \If{Using unitary model}
        \State $U_k \gets \text{proj}[U_k]$ using \eqref{eq:uni_projection}
    \EndIf
    \State $V_T \gets U_k V_T$
\EndFor
\State $U_K \gets V_0 V_T^{-1}$
\end{algorithmic}
\end{algorithm}

\section{Numerical simulation}\label{sect:simulation}
In this section, we analyze the efficiency of the proposed algorithm in the presence of various kinds of errors. In addition, we compare it with the optimization algorithm from \cite{kuzmin2021architecture}.

\subsection{Error models}\label{sect:error_model}
We define random mixing layers in two steps. In the first step, we generate a random unitary matrix
\begin{equation}
    U_H^\gamma = \proj[U_H + \gamma G], \quad
    G_{mn} \sim \Normal(0,1) + i\Normal(0,1), \quad
    m,n = 1,\dots,N,
\end{equation}
where $\proj[\cdot]$ is the unitary projection operation \eqref{eq:uni_projection}, and $\gamma$ characterizes the degree of distinction of $U_H^\gamma$ from the $N$-dimensional Hadamard transform, proportional to the discrete Fourier transform matrix  (the reconstruction method is not restricted to this particular type of mixing layers and could by applied to any invertable matrix):
\begin{equation}
    [U_H]_{mn} = \frac1{\sqrt{N}}\exp(-i\frac{2\pi(m-1)(n-1)}{N}), \quad m,n = 1,\dots,N.
\end{equation}
Further we assume $\gamma = 0.1$. In the second step, if necessary, we introduce heterogeneous linear losses:
\begin{equation}\label{eq:random_mixing}
    U_H^{\gamma,\beta}=\sqrt{U_H^\gamma}R_\beta\sqrt{U_H^\gamma}, \quad
    R_\beta = \diag[1-\beta r_n], \quad
    r_n \sim \Uniform(0,1), \quad
    n = 1,\dots,N.
\end{equation}
In other words, we insert random linear losses of no more than $\beta$ into the ``middle'' part of each random unitary transformation. Random matrices of the form \eqref{eq:random_mixing} are generated independently for each mixing layer to create a true model of the programmable interferometer. We denote these matrices as $U_k^{\text{true}}$.

We also simulate the errors of phase layers control. Let $\{\varphi_n\}$ be the desired set of phases in the $k$-th phase layer. The relative control error of each phase is given by $\alpha$:
\begin{equation}\label{eq:phase_error}
    \varphi_n^\alpha = \varphi_n(1+\alpha g_n), \quad
    g_n \sim \Normal(0,1), \quad
    n = 1,\dots,N.
\end{equation}
We simulate the cross-talk between the phases of a the single phase layer as following:
\begin{equation}
    \varphi_n^{\alpha, w} = \sum_{m=1}^{N}{\varphi_n^\alpha \exp(-\frac{(n-m)^2}{2w^2})}, \quad n=1,\dots,N,
\end{equation}
where $w$ characterizes the cross-talk strength. The errors of the type \eqref{eq:phase_error} are simulated independently for each phase layer each time the control phases of the interferometer are changed. An important condition is the absence of cross-talk between different phase layers. This condition is usually satisfied in the experiment, since the corresponding control elements (electrodes or heaters) are usually quite far apart on the optical chip.

We simulate the statistical noise of the transfer matrix measurements by a random Gaussian error for each matrix element:
\begin{equation}\label{eq:tomo_noise}
    V \rightarrow V + \varepsilon \sqrt{L} G, \quad
    G_{mn} \sim \Normal(0,1) + i\Normal(0,1), \quad
    m,n = 1,\dots,N,
\end{equation}
where $\varepsilon$ is the noise level, and $L$ is the measurement protocol size (the number of transfer matrices to measure). 
The $\varepsilon$ value includes all the statistical errors in the transfer matrix measurement procedure (e.~g. described in \cite{jacob2018direct}) and depends on the experiment details. 
The correction $\sqrt{L}$ is based on the standard assumption that statistical errors are inversely proportional to the square root of the acquisition time. It allows to compare different methods under the conditions of a fixed data acquisition time. For example, let Method~A requires to measure $50$ transfer matrices and Method~B requires $100$. Then, if the experiment duration is limited, one needs to reduce the measurement time twice for each matrix for Method~B (we neglect the phase layers configuration switching time). This relatively increases the variance of matrix elements estimates by a factor of $2$ for Method A. The linear error increases $\sqrt{2}$ times. The error \eqref{eq:tomo_noise} is generated independently each time, when the algorithms need to obtain a new transfer matrix.

\subsection{Benchmarks}
After the numerical experimental data generation we reconstruct the interferometer model. Then we estimate the ability of the model to predict the full transfer matrix given arbitrary phase layer configuration (each phase of each phase layer is chosen randomly from $0$ to $2\pi$) by calculating the fidelity between the predicted $V_{\text{pred}}$ and true $V_{\text{true}}$ transfer matrices of the entire interferometer (the normalized squared dot product of the matrices):
\begin{equation}\label{eq:benchmark_df}
    F = \frac{\abs{\Tr(V_{\text{true}}^\dagger V_{\text{pred}})}^2}{\Tr(V_{\text{true}}^\dagger V_{\text{true}})\Tr(V_{\text{pred}}^\dagger V_{\text{pred}})}.
\end{equation}
The quantity $\Delta F = 1 - F$, in turn, plays the role of the distance between the matrices, which is insensitive to the value of the undetectable global phase (similar to the distance between two quantum processes \cite{gilchrist2005distance}). We consider the $95$-th percentile $\Delta F_{95}$ in a test set of $1000$ random phase layer configurations \cite{bantysh2021quantum}. Note that the value of $\Delta F_{95}$ is affected not only by the quality of the mixing layers reconstruction, but also by the presence of phase layer control errors in $V_{\text{true}}$, which are assumed to be zero in $V_{\text{pred}}$.

In Section~\ref{sect:no_losses}, we pointed out that the method makes it possible to uniquely determine the normalized transmission coefficients \eqref{eq:transmission} of mixing layers. Therefore, we also consider the measure
\begin{equation}\label{eq:benchmark_dt}
    \Delta T_{\max} = \max_{\substack{k=1,\dots,K\\m,n=1,\dots,N}}{\abs{[T_k^{\text{true}}]_{mn} - [T_k^{\text{rec}}]_{mn}}}.
\end{equation}
It characterizes the maximal error of transmission coefficients estimation over the all mixing layers. Here $T_k^{\text{true}}$ and $T_k^{\text{rec}}$ are respectively the true and reconstructed transmission matrices of the $k$-th mixing layer. Both $\Delta F_{95}$ and $\Delta T_{\max}$ are insensitive to homogeneous losses. Considering $\Delta T_{\max}$ in addition to $\Delta F_{95}$ is useful for two reasons. First, $\Delta F_{95}$ is an integral estimate that includes both errors in determining mixing layers and errors in controlling phase layers. Since the latter remains unknown, $\Delta F_{95}$ may be quite high even if each mixing layer is determined accurately. Secondly, the determination of the transmission coefficient is of independent interest for debugging the manufacturing technology of integrated interferometers. The use of $\Delta T_{\max}$, however, makes sense for the unitary model only (see Section~\ref{sect:no_losses}). In the presence of heterogeneous losses, the matrices $T_k^{\text{rec}}$ are determined ambiguously and it makes no sense to compare them directly with $T_k^{\text{true}}$.

\subsection{Comparison with optimization method}
Let us consider the case when the number of interferometer channels $N$ and the number of mixing layers $K$ are 5 ($N = K = 5$), and there are no linear losses ($\beta=0$) and phase control errors ($\alpha=w=0$). We generate a random programmable interferometer according to Section~\ref{sect:error_model} and perform the reconstruction of all its mixing layers according to Section~\ref{sect:reconstruction}. As in \cite{kuzmin2021architecture}, we use the model of unitary mixing layers. The initial data used in the algorithm is also fed to the input of the algorithm from \cite{kuzmin2021architecture} with an open source code. We use the version of this algorithm that uses the approximate prior knowledge of mixing layers transfer matrices. Following the example from the library files, the following algorithm parameters are chosen: mini-batch size is five, the prior knowledge coefficient $\beta=0.24$. The number of training epochs varies from $20$ to $1500$ depending on the dimension $N$ and the measurement protocol (the value is high enough to reliably rich the accuracy saturation).

\begin{figure}[htbp]
\centering\includegraphics[width=\linewidth]{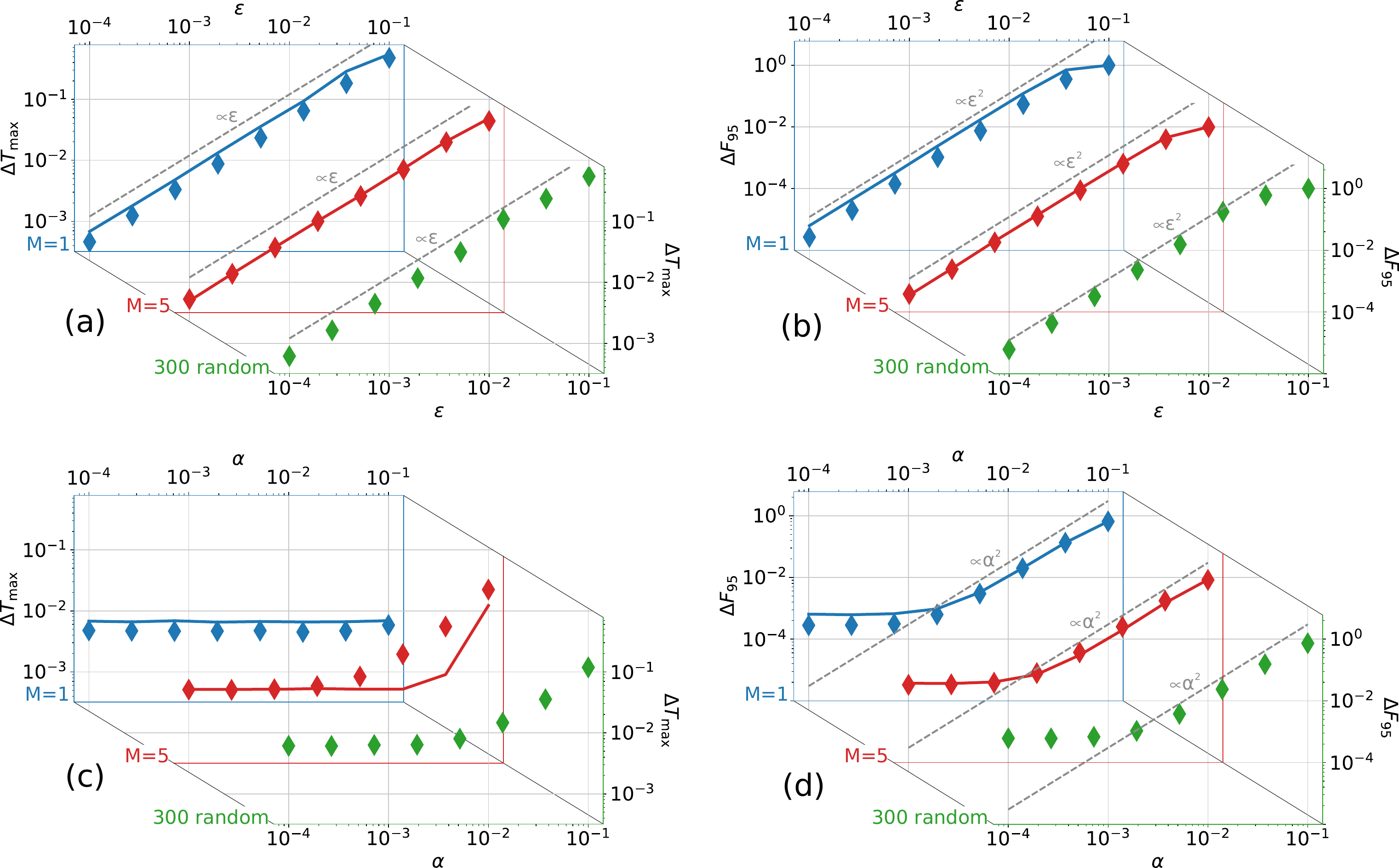}
\caption{The errors in determining the transmission coefficients $\Delta T_{\max}$ (a,c) and predicting the full transfer matrices $\Delta F_{95}$ (b,d). The average values over 100 independent trials are shown. Dimension $N=5$, number of mixing layers $K=5$, no losses ($\beta=0$). The results for our algorithm (solid lines) and for the algorithm from \cite{kuzmin2021architecture} (diamonds) are shown. (a,b) Dependence on the measurement error $\varepsilon$; no phase control errors ($\alpha=w=0$). (c,d) Dependence on the phase control error $\alpha$; measurement error $\varepsilon=10^{-3}$, cross-talk strength $w=0.1$.}
\label{fig:methods_comparison}
\end{figure}

In total, we considered 3 different measurement protocols:
\begin{enumerate}
    \item First, all phases are set to zero. Next, $20$ configurations are considered. In each one a single non-zero phase is set equal to $2\pi/3$. Each of these $20$ configurations allows one to define $M = 1$ column of the inverse transfer matrix of one mixing layer (see Section~\ref{sect:noise_handling}). The total number of configurations is $L = 21$.
    \item First, all phases are set to zero. Next, $4$ configurations are considered. In each one the phases \eqref{eq:phases_choice} are set on one of the phase layers. Each of these $4$ configurations allows one to determine all $M = 5$ columns of the inverse transfer matrix of one mixing layer (see Section~\ref{sect:noise_handling}). The total number of configurations is $L = 5$.
    \item A set of $L = 300$ random interferometer configurations is considered (only for the optimization algorithm from \cite{kuzmin2021architecture}).
\end{enumerate}
Fig.~\ref{fig:methods_comparison} shows the simulation results. In the absence of phase layers control errors, the dependencies of $\Delta T_{\max}$ (Fig.~\ref{fig:methods_comparison}a) and $\Delta F_{95}$ (Fig.~\ref{fig:methods_comparison}b) on the measurement error $\varepsilon$ turned out to be close, regardless of the measurement protocol and the reconstruction method used. The closeness is so high that plotting them in the same axes on a log scale make the distinction barely seen. Approximately, we have observed $\Delta T_{\max}\propto \varepsilon$ and $\Delta F_{95}\propto\varepsilon^2$. From Fig.~\ref{fig:methods_comparison}c it can be seen that the presence of phase layer control errors $\alpha$ does not affect our method's ability to estimate the mixing layer transmission coefficients $[T_k]_{mn}$: it is limited by the interferometer measurements accuracy only. This is because the algorithm uses the phases values only to do the correct sorting. However, starting from a certain critical value of the error $\alpha_c$, the accuracy of determining $[T_k]_{mn}$ sharply decreases, since the sorting is violated. The value of $\alpha_c$ can be increased by decreasing the number of unique phases at each iteration (it is defined by the parameter $M$ of the algorithm). The optimization algorithm from \cite{kuzmin2021architecture} relies on the particular phase values, so the control errors impair the estimation of $[T_k]_{mn}$. Despite this, the algorithm manages to train the entire interferometer as a whole with an error $\Delta F_{95}$ not higher than that provided by our algorithm (Fig.~\ref{fig:methods_comparison}d). Approximately, we have observed $\Delta F_{95}\propto\alpha^2$.

\begin{figure}[htbp]
\centering\includegraphics[width=.6\linewidth]{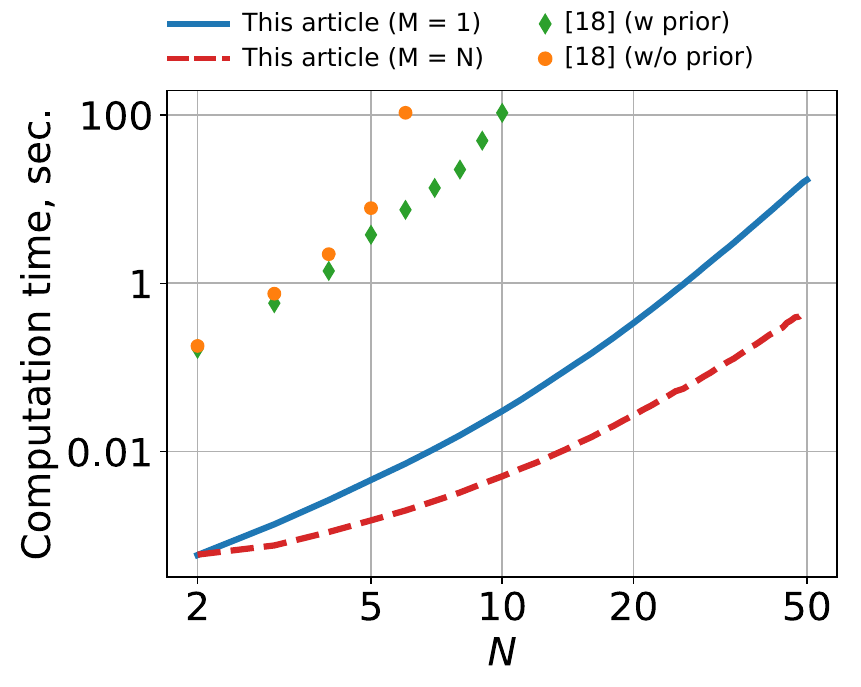}
\caption{Comparison of the computation time for different methods versus the dimension $N$. The average values over 100 independent trials are shown. The number of mixing layers $K = N$, no losses ($\beta = 0$), no phase control errors ($\alpha=w=0$), transfer matrices measurement error $\varepsilon=10^{-5}$.}
\label{fig:speed}
\end{figure}

Despite the closeness of the results for optimization algorithm and our algorithm, the latter performs much faster. To show this, consider two extreme versions of our algorithm ($M = N$ and $M = 1$). For the optimization algorithm from \cite{kuzmin2021architecture}, we use a set of 300 training phase layer configurations. Starting from a certain value, changing this number does not affect the computation speed significantly. The number of epochs is chosen to be minimal in order to guarantee a correct result. On the range of considered dimensions $N$, the presented reconstruction algorithm worked more than two orders of magnitude faster (Fig.~\ref{fig:speed}). We also considered the optimization algorithm in the ``black box'' model (the absence of any prior knowledge about the mixing layers). In this case, as expected, there was a rapid growth in computational complexity with an increase in $N$ from two to six (as in \cite{kuzmin2021architecture}, we also failed to obtain the convergence of the optimization algorithm in the ``black box'' model for $N > 6$). Our algorithm, also based on the ``black box'' model, showed a much slower increase in computation time. The independent trials of numerical experiments were performed in parallel on 52 CPU cluster; Intel(R) Xeon(R) CPU E5450 @ 3.00GHz.

\subsection{Algorithm scalability}
Above we have mentioned that the total number of required interferometer configurations for our method is $L=(K-1)\lceil N/M\rceil +1$, where $N$ is the number of input and output channels, $K$ is the number of mixing layers, and $M\in[1;N]$ is the number of simultaneously estimated columns of the inverse matrix of a single mixing layer. The time required to measure the transfer matrix for each configuration is estimated as $O(N^2)$ \cite{jacob2018direct}. Therefore, for $K=N$ the total measurement time scales as $O(N^4/M)$.

The reconstruction algorithm requires $O(L)$ algebraic operations including matrix multiplication, taking the inverse matrix, and computing matrix eigendecomposition. The complexity of these procedures does not exceed $O(N^3)$ \cite{demmel1997applied}. For $K=N$ the total complexity of the entire procedure can be estimated as $O(N^5/M)$.

\begin{figure}[htbp]
\centering\includegraphics[width=\linewidth]{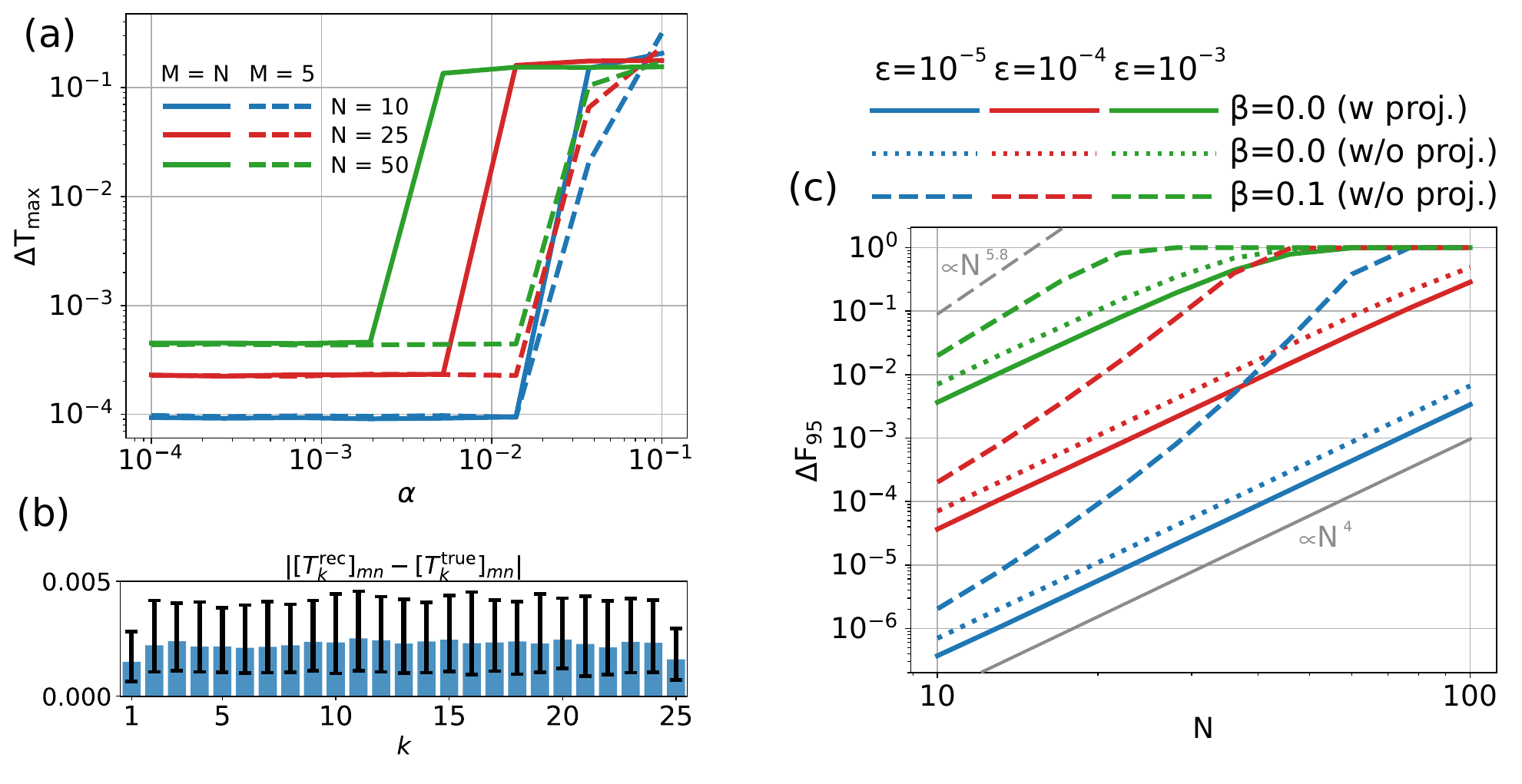}
\caption{(a) Scalability of the errors $\Delta T_{\max}$ in determining the mixing layers transmission coefficients. The average values over 100 independent trials are shown. Number of mixing layers $K = N$, no losses ($\beta = 0$), cross-talk strength $w=0.1$, transfer matrices measurement error $\varepsilon=10^{-5}$. (b) Distribution of transmission coefficients errors for various mixing layers within a single trial; confidence intervals show the upper and lower quartiles over all $m,n=1,\dots,N$. Parameters: $N=K=25$, $\beta=0$, $w=0.1$, $\alpha=10^{-3}$, $\varepsilon=10^{-5}$. (c) Scalability of the errors $\Delta F_{95}$ in predicting the full transfer matrices. The average values over 100 independent trials are shown. Parameters: $K = N$, $M = N$, $\alpha=w=0$.}
\label{fig:scalability}
\end{figure}

From these estimates, one can see the number $M$ to be an important parameter that affects the algorithm speed. The value $M = N$ is optimal from the point of view of measurement and computation speed. However, in this case, with increasing $N$, the sensitivity to control errors of phase layers increases: if the error $\alpha$ is higher than the critical value $\alpha_c$, the mixing layers estimates accuracy sharply decreases. At the same time, for a fixed $M$, the value of $\alpha_c$ changes slightly with increasing $N$ (Fig.~\ref{fig:scalability}a). Note also that for $\alpha<\alpha_c$ transmission coefficients matrices $T_k$ are determined with approximately the same accuracy for all mixing layers (Fig.~\ref{fig:scalability}b). That is, despite the iterative nature of the algorithm (see eq. \eqref{eq:ak_mat}), there is no effect of error accumulation.

The prediction error $\Delta F_{95}$ grows polynomially with increasing $N$ (Fig.~\ref{fig:scalability}c). This is primarily because at a fixed data acquisition time, the size $L$ of the measurement protocol increases. As a result, the measurement accuracy of each transfer matrix decreases (see eq. \eqref{eq:tomo_noise}), impairing the mixing layers estimation accuracy. The figure shows that one can slightly improve the accuracy by performing the projection of the mixing layers to the set of unitary matrices (Section~\ref{sect:no_losses}). This, however, does not affect the rate of growth (about $\Delta F_{95} \propto N^4$). When the heterogeneous linear losses are present, the rate of growth becomes higher (about $\Delta F_{95} \propto N^{5.8}$ for $\beta=0.1$). This can be explained by the fact that the condition number of the transfer matrices of the mixing layers increases when heterogeneous losses occur. As a result, the matrix inversion procedure becomes more sensitive to statistical fluctuations.

\section{Conclusion}
We have shown that the estimation of mixing layers in the layered architecture of the programmable integrated optical interferometer is possible without any use of numerical optimization. The proposed algorithm has a single parameter $M$ –- the number of simultaneously estimated columns of a mixing layer inverse transfer matrix. We have performed the numerical simulation and compared the results with the results obtained using the optimization algorithm. It has been found that under the equal conditions both methods provide the same accuracy, regardless of the measurement protocol. At the same time, the exclusively algebraic approach in our algorithm has shown a significantly lower computation time.

Moreover, our algorithm uses the control phases values only to correctly sort the columns of the adjacent mixing layer. This makes the method almost insensitive to phase control errors up to some critical error value. This critical value can be increased by choosing a smaller value of the parameter $M$. Small $M$, however, increases the number of operations in the reconstruction algorithm, so in practice it is worth choosing an optimal $M$ based on the experimental accuracy of phases control. We believe that extending the algorithm to consider the absolute values of the phases can help in determining the phase control errors. This is the subject of further research.

Finally, note that any interferometer architecture (for example, the Reck~\cite{reck1994experimental} or Clements~\cite{clements2016optimal} architecture) can be brought to the layered architecture considered here by adding controllable phase shifters in some channels. However, the possibility to directly apply our method to an arbitrary architecture remains unclear and is also the subject of further research.

\begin{backmatter}
\bmsection{Acknowledgments}
This work was supported by Russian Science Foundation (RSF), project No: 22-12-00263. We also thank I.~Dyakonov for valuable discussions and G.~Struchalin for help in conducting computations.

\bmsection{Disclosures}
The authors declare no conflicts of interest.

\bmsection{Data Availability}
Data underlying the results presented in this paper are not publicly available at this time but may be obtained from the authors upon reasonable request.
\end{backmatter}


\bibliography{sample}

\end{document}